# High-Energy Microresonator Soliton Generation

Zhenhua Guo, Sushant Kumar, Xue Dong, Yi Zhang, Jiewei Xiang, Arunima Nauriyal, Junchi Zhang, Elias Veilleux, Jaime Cardenas, and William H. Renninger

Institute of Optics, University of Rochester, Rochester, NY 14627, USA
*e-mail address: zguo14@ur.rochester.edu

## Abstract

Kerr resonators generate stable frequency combs in a compact platform with applications in coherent communications, sensing, quantum information processing, and astrophysics. Ultrashort pulses can be generated, moreover, with a wavelength and repetition rate flexibility inaccessible by traditional mode-locked lasers, which is desirable for high peak-power applications including in biomedicine and materials processing. However, for these applications, single-pulse energies will need to be significantly improved beyond the fJ level characteristic of sources today. Here we describe and demonstrate a simple approach for increasing the pulse energy in anomalous dispersion Kerr microresonators. Through scaling laws based on a mean-field cavity model and supported by experimentally-accurate numerical simulations, we show that pulse energy scales strongly with output coupling if supported by sufficient drive power. In strongly over-coupled cavities, femtosecond pulses can be stabilized with pulse energy beyond the pJ level. With this theoretical framework, by pumping a 12 GHz $Si_3N_4$ cavity with 30% output coupling with 2.6-ps time-lens generated pulses, we observe stable 74-fs pulses with a record output pulse energy of 6 pJ. High energy Kerr resonators are anticipated to improve performance for current applications and compliment mode locked lasers for high peak power applications.

## Introduction

Ultrashort pulses and frequency combs generated from Kerr microresonators are stabilized through a dissipative balance between loss, parametric four-wave mixing gain, dispersion, and nonlinearity[1]. Research to date has largely focused on applications that benefit from the bandwidth, coherence, or narrow linewidths of the generated frequency combs such as for coherent communication[2], RF frequency synthesis[3], self-referencing[4,5], distance ranging[6,7], and astrophysical spectrometer calibration[8]. Since these applications benefit from comb stability, high conversion efficiency, material diversity, and broad bandwidth, many significant advances have been developed in these areas. For example, for stability and managing thermal effects researchers have adopted power kicking[9], fast on-chip thermal tuning[10], and the use of auxiliary lasers[11]. For improving energy conversion efficiency, the primary approach is to increase the temporal overlap of the drive with the generated field to maximize the fraction of drive power that can convert. For example, higher order soliton crystal states can achieve an energy conversion efficiency around 50%[12], and normal dispersion cavities producing dark soliton combs can produce very high conversion efficiencies[13-15]. Additional improvements can be achieved by recycling some of the pump as with a photonic crystal reflector in[16]. For single soliton generation larger temporal overlap requires driving with short pulses where efficiencies can be raised from 1%[17] to 5%[18]. More recently, by further shortening the pump pulse and optimizing the external quality factor, single-pulse efficiencies as high as 34% have now been achieved[17]. Research has also focused on extending comb generation to a range of material platforms, including $MgF_2$[19], Si[20], and $Si_3N_4$[10,21]. Silicon nitride, for example, is a mature and versatile platform that supports low pump power thresholds due to its high nonlinearity. Finally, to achieve broader bandwidth combs, dispersion engineering is a primary technique. By tailoring the cavity waveguide geometry, the effective refractive index and thus the group velocity dispersion (GVD) can be controlled. Moving the GVD toward the zero-dispersion point can greatly broaden the comb bandwidth, which leads to supercontinuum generation[22].

However, despite the improvements described above, the demonstrated single-pulse energies remain below the pJ level, and an upper bound on achievable soliton pulse energy has not been clearly established or discussed. High-peak-power solitons are essential for bio-imaging[23], supercontinuum generation[24,25], material processing[26], precision surgery, and studies of nonlinear optical physics. Today these applications are serviced by mode-locked laser sources, which can readily provide kW peak powers, but are bulky, alignment constrained, and costly. This has motivated recent research in integrating mode-locked lasers on chip[27-31], with recent studies featuring pulses with 4.2 pJ[31] and 1nJ[32] energies, for example. However the former result featured longer pulses with 4.5 ps with lower peak power, and the latter, based on a Mamyshev oscillator technique, was not self-starting, requiring an external pulse to initialize and stabilize the process. While certainly promising, as elaborated in the discussion, integrated gain is challenging to implement, and ultimately, will limit the possible soliton wavelengths, the cavity repetition rate, and can introduce undesirable carrier effects, noise and nonlinearities. Therefore, because the Kerr micro resonator platform does not require a gain section, extending its soliton pulse energy into the high-power regime could allow micro resonators to begin supplanting mode-locked lasers in ultrashort-pulse applications such as bio-imaging and materials processing. Higher pulse energy would also strengthen conventional frequency-comb applications by enabling, for example, longer-range operation in coherent LiDAR and by providing higher per-line power for parallel coherent optical communication.

In this work, we investigate the conditions for generating high energy single solitons from anomalous dispersion Kerr resonators. Beginning with simple scaling laws from the approximate Lugiato–Lefever equation (LLE) model, we find that high energy solitons can be accessed in highly over-coupled cavities given sufficiently high pump powers. Full experimentally accurate numerical simulations validate this conclusion and provide practical design guidelines. Experimentally, high drive powers are achieved through electrically generated picosecond pulses and on-chip mode converters for high fiber-to-chip coupling efficiency, and high pump pulse energies are achieved by operating with low, 12 GHz, repetition rate. While driving $Si_3N_4$ microresonators in the forward direction, a counter-propagating auxiliary laser is used for thermal stabilization. In a cavity with 30 % output coupling, we obtain stable 74-fs pulses with 6 pJ output pulse energy and validate coherence with autocorrelation measurements in agreement with predictions. To our knowledge this represents the highest on-chip Kerr-soliton pulse energy and the highest on-chip self-starting pulse energy reported to date. This result and the associated design guidelines motivate further development of high-energy micro resonator sources for ultrashort-pulse applications.

## Analytical theory

In the limit of small changes to the pulse in the cavity, the approximate mean-field model of the cavity described by the generalized mean-field Lugiato–Lefever equation can provide insight into how the output-pulse energy scales with cavity parameters such as loss, length, detuning, drive, nonlinearity and dispersion. Following previous treatments[33], the LLE can be rewritten so that distance (z), time (T), and power (P) are normalized as:

$$\frac{z}{z_{n0}} = \frac{1}{L}, \qquad \frac{T}{T_{n0}} = \sqrt{\frac{L|\beta_2|}{\alpha}}, \qquad \text{and} \qquad \frac{P}{P_{n0}} = \frac{\alpha}{\gamma L}, \qquad (1)^{33}$$

where $\alpha$ is the total round-trip loss (bus-to-ring coupling $\alpha_c$ plus cavity intrinsic loss $\alpha_i$), $L$ is the cavity length, $\beta_2$ is the group velocity dispersion, $\gamma$ is the nonlinear coefficient, and $z_{n0}$, $T_{n0}$, and $P_{n0}$ are the normalized distance, time, and powers, respectively. It follows that energy is normalized by $\frac{E}{E_{n0}} = \frac{T}{T_{n0}} * \frac{P}{P_{n0}} = \frac{1}{\gamma}\sqrt{\frac{\alpha|\beta_2|}{L}}$. The LLE can then be written in terms of two normalized parameters:

$$D_{n0} = \frac{D\gamma L}{\alpha^3} \qquad \text{and} \qquad \delta_{n0} = \frac{\delta}{\alpha}, \qquad (2)$$

where $D$ is the intracavity drive power and $\delta$ is the pump–cavity detuning. If a single-soliton solution exists for the set $(z_{n0}, T_{n0}, P_{n0}, E_{n0}, D_{n0}, \delta_{n0})$, then any set of $(z, T, P, E, D, \delta)$, that satisfies Eq. (1-2) also supports a soliton. This leads to the following scaling laws:

$$D \propto \frac{\alpha^3}{\gamma L}, \qquad \delta \propto \alpha, \qquad \Delta T \propto \sqrt{\frac{L|\beta_2|}{\alpha}}, \qquad P \propto \frac{\alpha}{\gamma L}, \quad \text{and} \quad E \propto \frac{1}{\gamma}\sqrt{\frac{\alpha|\beta_2|}{L}} \qquad (3)^{33}$$

Here we see that intra-cavity pulse energy scales with nonlinearity, loss, dispersion and cavity length. Nonlinearity is mostly fixed by the cavity material and while energy increases with dispersion, the pulse duration does as well so this does not bring a net advantage for peak power. Energy has a weak inverse dependence on length, and when the length decreases the pulse duration does as well, which is desirable for peak power. However, a shorter length means lower energy for a fixed average power and since this available energy would decrease faster than the soliton energy increases with a decrease in length, this variable will have less impact. Finally, intra-cavity energy is proportional to the square root of the total cavity loss, which leads to a clear recipe for improvement. The intracavity energy scales as $\alpha^{0.5}$, but because the energy extracted from the cavity is linearly proportional to the coupling coefficient, the output energy scales as $\alpha_c \alpha^{0.5}$. If the chip is strongly over-coupled, the intrinsic loss can be approximated as negligible and $\alpha \approx \alpha_c$. This means that the extracted soliton energy scales as $\alpha^{1.5}$, where $\alpha$ is the output coupling. The intracavity drive required to sustain the soliton scales as $D \propto \alpha^3$. After accounting for coupling losses, the required external pump power $P_{\text{ext}}$ therefore grows approximately proportional to $\alpha^2$. If we further assume that the required pump pulse duration is scaled proportionally with the generated soliton pulse duration, then the required external pump energy exhibits the same $\alpha^{1.5}$ dependence. (Fig. 1) This scaling implies that the energy conversion efficiency remains approximately constant across different pump-energy operating conditions. However, this does not imply that soliton energy can be increased without bound by indefinitely increasing the coupling fraction and drive power. When the intracavity field undergoes sufficiently large variation over a round trip, the mean-field approximation underlying the LLE is no longer valid, and the model predictions do not hold. Before this limit though, the LLE analysis shows that devices with stronger coupling should have the capability to support stable Kerr solitons with higher pulse energy. Intuitively, a larger bus-to-ring coupling fraction increases the round-trip loss and reduces the effective intracavity nonlinear interaction per round trip. At the same time, stronger coupling relaxes the intracavity power constraint and improves soliton extraction efficiency to the bus waveguide. As a result, higher intracavity peak powers and larger soliton energies can still be achieved, provided that the external pump power and pulse energy are scaled accordingly.

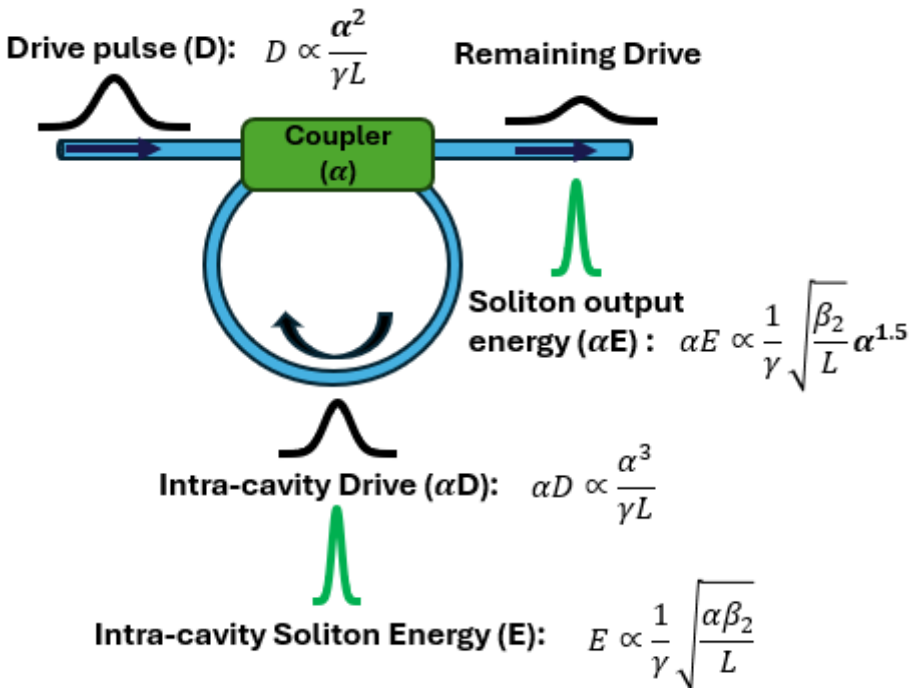

Figure 1: Approximate soliton energy scaling laws in anomalous dispersion resonators.

## Cavity design

Kerr-comb generation requires a device platform that simultaneously provides low optical loss and sufficient nonlinearity. Compared to alternatives such as silicon on insulator, lithium niobate, or silica, silicon nitride was selected due to its combination of relatively high nonlinearity, low material loss, low multi-photon absorption, and CMOS compatibility. We further select a thick waveguide design to maintain low propagation loss while enabling dispersion engineering to achieve suitable broadband anomalous dispersion for soliton generation. The waveguide cross-section was swept using PhotonDesign FIMMWAVE, and a geometry with approximately 780 nm thickness and 1.7 μm width was selected. With this design, we expect to achieve a cavity group-velocity dispersion of -150 fs^2/mm, a nonlinear coefficient of approximately 0.7 $(W\cdot m)^{-1}$, and an intrinsic quality factor on the order of 4~6 million.

Most previous on-chip soliton devices have low output coupling, for which a point coupler was sufficient for broadband pulse generation[1]. However, with over-coupled devices required for high energies, broadband coupling is more readily achieved with an asymmetric co-propagating coupler design, which has been shown to provide broadband coupling while also enabling systematic control of the coupling fraction through the waveguide gap and the interaction length of the coupler section[34]. By varying the coupling length of the coupler, we designed different bus-to-ring coupling percentages ranging from 2% to 50%.

The fabrication process and advanced mode converter design details are described in detail in the Methods section. Fabricated devices were characterized by frequency sweeping a laser through the cavity while simultaneously calibrating the laser with a fiber Mach–Zehnder interferometer [35], from which the intrinsic loss, external coupling strength, cavity FSR, and GVD can be extracted (see details in Supplementary Information)[19,36]. At 1550nm, the round-trip power loss is found to be ~2% (corresponding to a propagation loss of 0.074 dB/cm and an intrinsic quality factor of 5 million), the FSR is $12.2 \pm 0.01\ GHz$, the group index is $n_g = 2.11 \pm 0.002$, and the GVD is $\beta_2 = -(2 \pm 0.18) \times 10^5\ fs^2/m$.

## Numerical simulations

Using the measured cavity design parameters, numerical simulations can be used to accurately predict soliton pulse energy performance without approximation. The simulations are based on an Ikeda-type model[37], which accurately represents the Kerr cavity by describing propagation in the waveguide with the nonlinear Schrödinger equation while incorporating drive and loss at a distinct spatial location in the cavity. After fixing dispersion, nonlinearity, and loss to the designed and measured values, the remaining variables are coupling, drive power, and detuning. To evaluate the dependence on coupling, 6 bus-to-ring coupling values from 1% to 50% are examined, corresponding to the coupling configurations of the fabricated devices. For each coupler, simulations are run over a wide range of continuous wave drive and detuning values to generate an energy colormap of converged single-pulse solutions, where convergence is determined by verifying that intracavity peak power at each round trip reaches a constant value, and single-soliton solutions were identified using a peak-search procedure. To increase the excitation rate of single-soliton solutions, after initially seeding the simulations with a 1 ps Gaussian pulse, the resulting single-soliton solution was then used as the initial condition to seed nearby parameter regimes. Each stable single-soliton point is color-mapped according to its energy (a,d), efficiency (b,e), and roundtrip B-integral (c,f) and other field profiles, including non-converged simulations, multi-soliton states, and continuous-wave solutions, are represented by white regions.

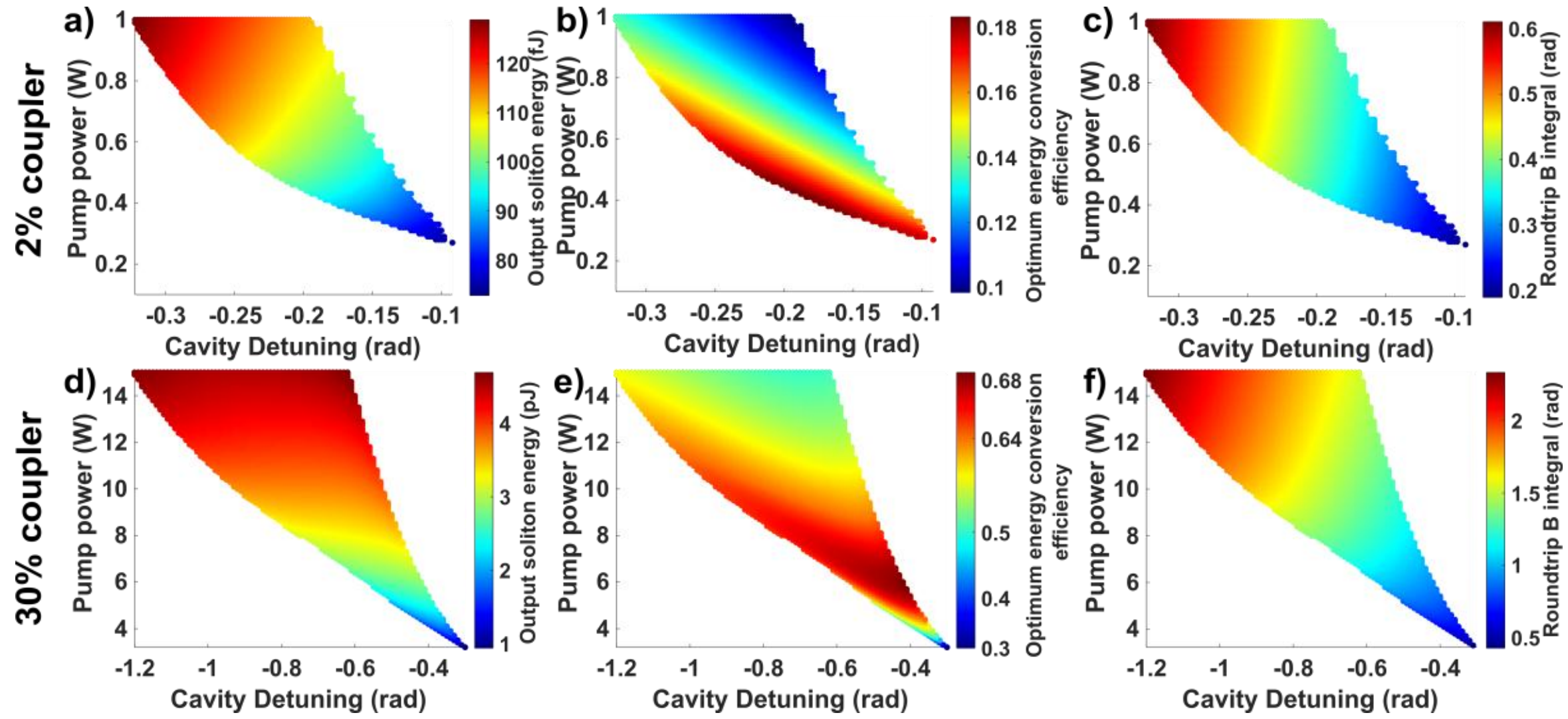

Figure 2: (a,d) Output soliton energy, (b,e) optimum energy conversion efficiency and (c,f) intracavity B-integral for the (a-c) 2 % and (d-f) 30 % coupler device.

We begin by examining specific couplers chosen to highlight two different design philosophies (Fig. 2). In the typical case, with a 2% coupler, the cavity is critically coupled to provide the lowest pump requirement. In contrast, with a 30% coupler, high energy solitons can be generated. Output soliton energy, pump-to-soliton energy conversion efficiency, and round-trip nonlinearity (B-integral) are examined to distinguish between these regimes. The optimum energy conversion efficiency was defined as the ratio of the soliton output pulse energy to the minimum required pump energy as determined by the drive power integrated over the time duration of the interaction (described in Methods).

A notch filter with a bandwidth corresponding to this interaction time is applied to remove the pump, as in the experiment. The output soliton energy is obtained after application of this filter. Finally, the round-trip B-integral is calculated by integrating the intracavity peak power over the cavity propagation length and multiplying by the nonlinear coefficient of the cavity. More details are discussed in the supplementary. While the 30% coupler device requires much higher pump power than the 2% coupler device, it can also generate substantially higher soliton pulse energy, in good agreement with the analytical analysis. In addition, 30% coupling design enables higher pump-to-soliton energy conversion efficiency and greater tolerance to intracavity nonlinearity.

We can examine the dependence of energy, conversion efficiency, and accumulated nonlinearity vs coupling and drive power more broadly (Fig. 3), by fixing detuning to a value determined by Eq. 2. Interpolation was used for values between the seven simulated coupler values. The LLE scaling laws from Eq 3 (black solid lines) are plotted for reference showing that drive and alpha have a cubic relationship. For example, while the 2% coupler case has a stable soliton at an intracavity drive power of 1 mW, the 10% coupler case increases the drive to 125 mW. Also the 2% coupler case has a detuning of -0.1 rad, and the 10% coupler case scales linearly with $\alpha$, leading to -0.5 rad. The top (purple) and bottom (green) lines correspond to the highest and lowest drive for which soliton solutions were obtained. The top group corresponds to the highest-energy solitons, exhibiting the greatest tolerance to nonlinearity, while the bottom group gives the minimum drive power required to generate a stable soliton. Because in the high-nonlinearity and larger coupling regimes the round-trip variation of the intracavity optical-field intensity becomes sufficiently large that the mean-field approximation underlying the LLE is no longer valid, the upper limit does not follow the cubic scaling law. Higher coupling percentages can withstand slightly larger nonlinear phase shift (Fig. 3(c)), as their higher intracavity loss reduces the strength of the nonlinear interaction. The maximum tolerable nonlinear phase shift is approximately $2.1\pi$ at 30 %~50 % devices, which is close to the typical nonlinear phase shift tolerated in standard fiber mode-locked lasers[38]. The use of a strongly over-coupled device with appropriate drive power enables generation of solitons with output energies exceeding 10 pJ (Fig. 3(a)), representing a 100-fold increase compared to the typical 100 fJ energy level of conventional frequency combs. Different coupler devices exhibit an optimum efficiency at a specific drive power when the pump pulse duration is not restricted (Fig. 3(b)). The highest energy conversion efficiency, exceeding 60%, is achieved with the 20% and 30% coupler devices.

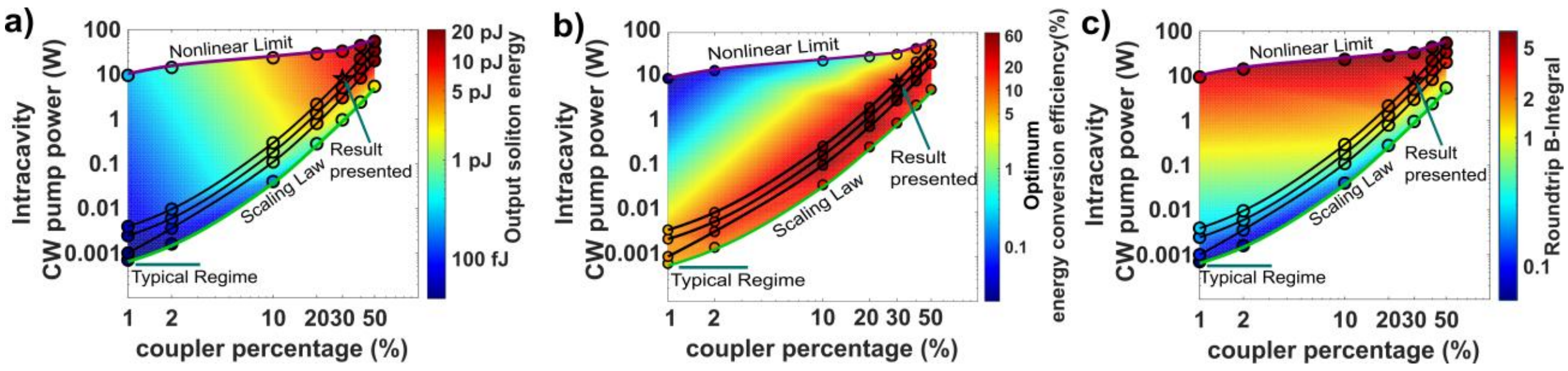

Figure 3. Log-scale colormap of (a) output soliton energy, (b) optimum energy conversion efficiency, and (c) roundtrip B-integral as a function of coupler percentage and pump power. The top boundary marks the nonlinear upper limit for stable-soliton existence, and the bottom boundary marks the lowest pump level at which stable solitons exist.

The analytical analysis and numerical simulations so far show that high-energy pulse generation requires a strong coupler design together with a high pump energy. At the same time, the overall performance must be evaluated in the context of energy conversion efficiency and practical limitations of the available pulse-pump system. Measurements of our fabricated devices identified the 30% coupler as the highest coupling device that still preserves a sufficiently high quality factor. Our pulse pump system can generate 2.6 ps pulses with up to 2 W average power after EDFA amplification, corresponding to 57 W peak power and 164 pJ pulse energy (Details in the supplement). After the 33% fiber-to-chip coupling efficiency and additional component losses, this yields about 40 pJ on chip pulse energy, or 14 W peak power. Since several fiber components have a 1.5 W average power damage threshold, we set the experimental target to 30 pJ on-chip pulse energy, corresponding to 10 W peak power, to maintain a sufficient margin. Based on these constraints, time lens simulations were performed to design a suitable pump pulse, as detailed in the Supplement. This simulated pulse was then used as the pulsed pump source in additional simulations refining the CW simulations presented above. The results indicate that this pump condition is expected to generate solitons with approximately 20% energy conversion efficiency, corresponding to a soliton pulse energy of 6 pJ for the 30% coupler devices.

# Experimental results

The experimental setup (Fig. 4a) begins with a time-lens pulse pumping system comprising a 1550 nm CW tunable laser, two low-$V\pi$ phase modulators, and one intensity modulator, generating 2.6 ps pulses at 12.2 GHz, matched to the cavity's free spectral range (FSR) (see Supplement for more details). The pulses are then amplified by an EDFA to 1.3 W average power, with 367 mW on chip after accounting for insertion and coupling losses, corresponding to a 10.4 W pulse peak power on chip. In parallel, a second CW tunable laser is amplified to ~2 W and injected in the counter-propagating direction as an auxiliary laser, passively stabilizing intracavity power drift and expanding the soliton access region[11]. The detailed experimental procedure to generate a single soliton state is described in the Methods. We use an optical spectrum analyzer to verify the formation of the single-soliton state, which is identified by a clean $\text{sech}^2$-shaped comb spectrum (Figure. 4(c)). The measured on-chip soliton power is 73 mW, corresponding to a pulse energy of 6 pJ. The on-chip pump pulse energy is 30 pJ, yielding a 20 % energy conversion efficiency. These measurements are in good agreement with the numerical model.

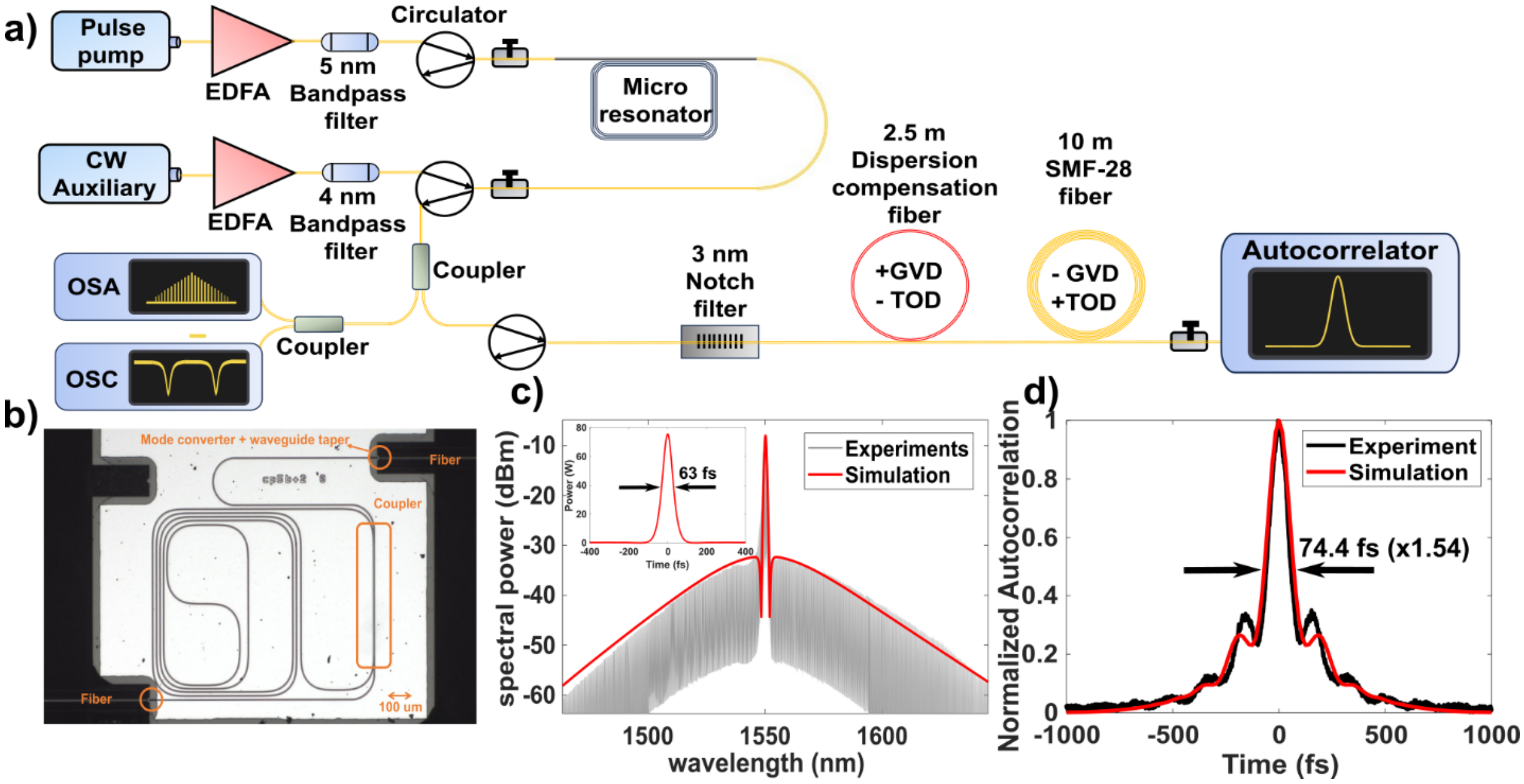

Figure 4: High energy Kerr soliton experiments. (a): Experimental setup for high-energy soliton generation. (b): Microscope image of the Si3N4 micro resonator with 30% output coupler. (c): Measured 6 pJ soliton spectrum with the numerical simulation envelope overlaid for comparison. Inset: Simulated temporal pulse shape of the 6 pJ soliton with 63 fs FWHM pulse duration. (d): Comparison between the shortest simulated and measured autocorrelation profiles.

For temporal characterization, the residual pump pulse is first removed with a 3.1-nm notch fiber Bragg grating (FBG) filter with 40 dB reflectivity. Dispersion-compensating fiber is then added to remove excess phase from the SMF-28 fiber following the chip. The resultant dechirped soliton autocorrelation is measured with a pulse duration of 74.4 fs. The observed secondary temporal structures (Fig. 4d)) are an artifact of the experimental setup and were found to arise from residual uncompensated spectral phase introduced by the FBG filter, combined with approximately 5% uncompensated TOD, as validated by numerical simulations using direct measurements of the FBG filter [39]. The details are explained in the supplementary.

## Discussion

Further increasing Kerr soliton pulse energy requires understanding what ultimately limits this platform. Here, the dominant constraint is round-trip nonlinearity, which reaches approximately 2π (Fig. 3(c)); beyond this, the soliton state becomes unstable and breaks up. Assuming a 50% coupler device can be realized while maintaining sufficient quality factor, this constraint sets a maximum soliton energy of approximately 20 pJ for the current platform. Extending the tolerable nonlinearity to still-higher pulse energies will require alternative designs, such as chirped-pulse Kerr resonators[33], in which the soliton is generated as a longer, chirped pulse: peak power is reduced while pulse energy is maintained through the increased duration.

Compared with on-chip mode-locked lasers, our high-energy-enabled Kerr resonator offers several advantages. First, pulse energy is no longer an exclusive advantage of mode-locked lasers: our 6 pJ, 74.4 fs on-chip Kerr soliton (71 W peak power) already exceeds the 4.2 pJ, 4.5 ps pulses (0.88 W peak power) typical of on-chip mode-locked lasers at comparable repetition rates[31]. Second, mode-locked lasers are fundamentally restricted to the bandwidth of their gain medium; producing powerful pulses outside the conventional Erbium- or Ytterbium-doped bands requires expensive, complex wavelength conversion. Kerr solitons face no such limitation, since they draw on parametric gain from four-wave mixing rather than a fixed gain medium[19]. Third, the gain sections themselves are difficult to design and fabricate within the compact, tightly integrated footprints on-chip lasers require, which in turn constrains the repetition rates they can reach. Fourth, gain dynamics, carrier effects, and amplified spontaneous emission (ASE) in the gain section degrade output pulse quality: carrier depletion produces time-dependent gain saturation and pulse-to-pulse energy and phase noise, ultrafast carrier effects reduce temporal coherence, and ASE lowers the optical signal-to-noise ratio and degrades comb-line phase coherence. Extending Kerr microresonator soliton energy into the high-power regime could therefore let microresonators begin supplanting mode-locked lasers for selected ultrashort-pulse applications such as bio-imaging and materials processing. For example, a particularly compelling application for this platform is soliton generation at new hard-to-reach wavelengths, such as near 1300 nm for biomedical imaging. Given a suitable pulse pump at 1300 nm, the microresonator can generate sub-ps solitons with kW-level peak power, sufficient for applications such as three-photon nonlinear spectroscopy[40,41].

For traditional Kerr-comb applications such as coherent LiDAR and WDM communications, a tenfold increase in soliton pulse energy translates directly into a tenfold increase in comb-line power at each wavelength, assuming comparable repetition rate and spectral envelope. This higher per-line power extends detectable range by increasing transmitted power and receiver SNR, and facilitates multi-channel and multi-beam architectures by leaving sufficient power per channel after demultiplexing and beam splitting; it also provides greater margin to compensate insertion losses from additional components such as spectral shaping or intensity modulation stages[42-45].

## Conclusion

In summary, we describe a pathway to high-energy soliton generation in anomalous-dispersion Kerr resonators, guided by approximate scaling laws and validated with numerical simulations. We showed that high output soliton energy can be achieved in devices with large output coupling driven with sufficient pump energy, enabling efficient energy extraction while maintaining tolerable intracavity nonlinear effects. Guided by this design principle, we measure 74.4 fs FWHM pulse duration solitons with record-level 6 pJ energy in a 12-GHz silicon nitride resonator with 30% coupling, experimentally establishing the approach. High-energy Kerr solitons are expected to improve the performance of existing microresonator-based applications and to complement conventional mode-locked lasers in high-peak-power photonic systems.

## Methods

### 1. Fabrication detail

Devices were fabricated on 4-inch silicon wafers with a 4 μm thermal $SiO_2$ undercladding. A 770 nm $Si_3N_4$ layer was deposited by low-pressure chemical vapor deposition (LPCVD) using the twist-and-grow process[46], rotating the wafer by 45° after every 300 nm of deposition; this achieves results comparable to the damascene reflow[47] and multi-step annealing[48] approaches while remaining compatible with our existing fabrication flow. A silicon oxide hard mask was then deposited by plasma-enhanced CVD (PECVD), patterned with MAN-2405 photoresist using a JEOL 9500 e-beam lithography system (multi-pass exposure), and etched in an Oxford 100 inductively coupled plasma (ICP) etcher. After stripping the photoresist with oxygen plasma cleaning, the silicon nitride waveguides were etched in the same ICP etcher; this oxide hard mask together with the high-selectivity etch chemistry reduces sidewall roughness and, in turn, waveguide propagation loss[49].

The waveguides were then clad with silicon oxide: 500 nm deposited by LPCVD, followed by an additional 3.5 μm by PECVD. The higher deposition temperature of the LPCVD step fully fills the high-aspect-ratio gap between the bus and ring waveguides, which can remain partially unfilled with PECVD oxide alone. Oxide mode converters (MCs) and fiber grooves were then patterned by deep-UV lithography (ASML PAS 5500) and etched in the Oxford 100 ICP etcher, followed by approximately 70 μm deep silicon grooves etched by deep reactive-ion etching (DRIE) in a Plasmatherm Versaline deep silicon etcher. After removing residual photoresist with oxygen plasma cleaning, a silicon undercut was performed with xenon difluoride ($XeF_2$) gas — critical for removing silicon beneath the mode-converter region and optically isolating the guided mode at the PIC facet. Chips were finally singulated on a DISCO dicing saw.

**2. Mode converter design**

A key limiting factor for high-power on-chip nonlinear phenomena is the optical power that can be coupled onto and off the photonic integrated circuit (PIC) — particularly for pulse pumping, where a pulse of specified spectrum, peak power, polarization, and duration must be delivered with high efficiency. The common approach, lensed-fiber coupling, is generally limited at high power by the small mode size and resulting high power density at the chip facet and taper region[50]. Expanding the effective mode size with an index-matched epoxy or fluid at the fiber–chip interface can reduce Fresnel back-reflection and improve coupling efficiency, but the polymer-based index-matching medium degrades under sustained high optical power and can ultimately cause damage at higher input levels. We instead designed silicon-oxide mode converters (MCs) for efficient fiber-to-PIC coupling, following the design procedure reported previously[51,52]. Because our platform uses thick silicon nitride, the coupling transition from the silicon nitride waveguide taper to the oxide MC could only be optimized to approximately 65% efficiency, compared with the >90% efficiency of prior demonstrations using 300 nm silicon nitride platforms.

**3. Extraction of the optimum energy conversion efficiency**

The optimum energy conversion efficiency is the ratio of the soliton output pulse energy to the minimum required pump energy, defined as the CW drive power multiplied by the minimum pump–soliton interaction time. To determine this interaction time, we calculate the change in intracavity field intensity over one round trip and take the absolute value of its derivative, which removes the slowly varying CW background from intrinsic cavity loss that is unrelated to the pump–soliton interaction. The optimum interaction time is then the interval over which this absolute derivative remains above 3% of its maximum value — below this threshold, the pump field is treated as effectively non-interacting with the soliton. The 3% threshold is chosen to match pulse-pumped simulations run at different pulse durations.

**4. Experimental procedure to generate soliton**

To locate the cavity resonance, the pump-pulse wavelength is swept periodically from blue- to red-detuned, with the auxiliary CW laser initially positioned outside resonance; at this stage the oscilloscope shows the pump resonance's thermally distorted, asymmetric triangular line shape. Tuning the auxiliary laser into resonance from the blue- toward the red-detuned side progressively narrows and symmetrizes this line shape back toward its original Lorentzian form. We stop the auxiliary laser near the resonance dip on its blue-detuned side — the condition that best compensates the pump-induced thermal effect — then finely tune the pump-pulse repetition rate near the measured cavity FSR until soliton steps appear in the red-detuned region. Stopping the pump wavelength sweep at one of these steps generates a soliton; the detuning can then be adjusted further to access single- or multi-soliton states.

## Reference

1. Kippenberg, T. J., Gaeta, A. L., Lipson, M. & Gorodetsky, M. L. Dissipative Kerr solitons in optical microresonators. *Science (80-. ).* **361**, (2018).
2. Marin-Palomo, P. *et al.* Microresonator-based solitons for massively parallel coherent optical communications. *Nature* **546**, 274–279 (2017).
3. Spencer, D. T. *et al.* An optical-frequency synthesizer using integrated photonics. *Nature* **557**, 81–85 (2018).
4. Jost, J. D. *et al.* Counting the cycles of light using an optical microresonator. *Opt. InfoBase Conf. Pap.* **2**, (2015).
5. Brasch, V., Lucas, E., Jost, J. D., Geiselmann, M. & Kippenberg, T. J. Self-referenced photonic chip soliton Kerr frequency comb. *Light Sci. Appl.* **6**, e16202-6 (2017).
6. Trocha, P. *et al.* Ultrafast optical ranging using microresonator soliton frequency combs. *Science (80-. ).* **359**, 887–891 (2018).
7. Suh, M. G. & Vahala, K. J. Soliton microcomb range measurement. *Science (80-. ).* **359**, 884–887 (2018).
8. Obrzud, E. *et al.* A microphotonic astrocomb. *Nat. Photonics* **13**, 31–35 (2019).
9. Brasch, V., Geiselmann, M., Pfeiffer, M. H. P. & Kippenberg, T. J. Bringing short-lived dissipative Kerr soliton states in microresonators into a steady state. *Opt. Express* **24**, 29312 (2016).
10. Joshi, C. *et al.* Thermally controlled comb generation and soliton modelocking in microresonators. *Opt. Lett.* **41**, 2565 (2016).
11. Zhang, S. *et al.* Sub-milliwatt-level microresonator solitons with extended access range using an auxiliary laser. *Optica* **6**, 206 (2019).
12. Ali Afridi, A. *et al.* Versatile power-efficient octave-spanning soliton crystals with high conversion in a Si3N4 microresonator. *CLEO Fundam. Sci. CLEOFS 2023* **31**, 33191–33199 (2023).

13. Macnaughtan, M., Erkintalo, M., Coen, S., Murdoch, S. & Xu, Y. Temporal characteristics of stationary switching waves in a normal dispersion pulsed-pump fiber cavity. *Opt. Lett.* **48**, 4097 (2023).

14. Bao, C. *et al.* Observation of Breathing Dark Pulses in Normal Dispersion Optical Microresonators. *Phys. Rev. Lett.* **121**, 257401 (2018).

15. Im, B. O. K. Y. O. K. *et al.* Turn-key , high-efficiency Kerr comb source. **44**, 2–5 (2019).

16. Godey, C., Balakireva, I. V., Coillet, A. & Chembo, Y. K. Stability analysis of the spatiotemporal Lugiato-Lefever model for Kerr optical frequency combs in the anomalous and normal dispersion regimes. *Phys. Rev. A - At. Mol. Opt. Phys.* **89**, (2014).

17. Li, J. *et al.* Efficiency of pulse pumped soliton microcombs. *Optica* **9**, 231 (2022).

18. Obrzud, E., Lecomte, S. & Herr, T. Temporal solitons in microresonators driven by optical pulses. *Nat. Photonics* **11**, 600–607 (2017).

19. Herr, T. *et al.* Temporal solitons in optical microresonators. *Nat. Photonics* **8**, 145–152 (2014).

20. Yu, M., Okawachi, Y., Griffith, A. G., Lipson, M. & Gaeta, A. L. Modelocked Mid-Infrared Frequency Combs in a Silicon Microresonator. *Opt. InfoBase Conf. Pap.* **3**, (2016).

21. Brasch, V. *et al.* Photonic chip based optical frequency comb using soliton induced cherenkov radiation. *Opt. InfoBase Conf. Pap.* 357–360 (2015).

22. Anderson, M. H. *et al.* Photonic chip-based resonant supercontinuum via pulse-driven Kerr microresonator solitons. *Optica* **8**, 771 (2021).

23. Fan, S., Wang, S., Yang, C., Wise, F. & Kong, L. Advances of Mode-Locking Fiber Lasers in Neural Imaging. *Adv. Opt. Mater.* **11**, 1–8 (2023).

24. Takushima, Y., Futami, F. & Kikuchi, K. Generation of over 140-nm-wide super-continuum from a normal dispersion fiber by using a mode-locked semiconductor laser source. *IEEE Photonics Technol. Lett.* **10**, 1560–1562 (1998).

25. Luo, X. *et al.* All-Fiber Supercontinuum Source Pumped by Noise-Like Pulse Mode Locked Laser. *IEEE Photonics Technol. Lett.* **31**, 1225–1228 (2019).

26. Kauf, M., Patel, R. & Bovatsek, J. High Speed Processing for Microelectronics and Photovoltaics. *Laser Tech. J.* **6**, 33–37 (2009).

27. Sohler, W. Harmonically mode-locked Ti : Er : LiNbO 3 waveguide laser. **20**, 596–598 (1995).

28. Guo, Q. *et al.* Ultrafast mode-locked laser in nanophotonic lithium niobate. **713**, 708–713 (2023).

29. Gasse, K. Van, Kjellman, J. Ø., Caër, C. & Nakamura, T. High-pulse-energy III-V-on-silicon-nitride. (2022) doi:10.1063/5.0058022.

30. Ling, J. *et al.* Electrically empowered microcomb laser. 1–8 (2024) doi:10.1038/s41467-

024-48544-2.

31. High-power heterogeneously integrated mode-locked laser enabled by a booster amplifier. **33**, 54747–54756 (2025).

32. Qiu, Z. *et al.* High-pulse-energy integrated mode-locked lasers based on a Mamyshev oscillator.

33. Spiess, C., Yang, Q., Dong, X., Bucklew, V. G. & Renninger, W. H. Chirped dissipative solitons in driven optical resonators. *Optica* **8**, 861 (2021).

34. Zhang, Y. & Cardenas, J. Integrated Silicon Nitride Devices for Novel Photonic Applications. (2023).

35. Inga, M. *et al.* Alumina coating for dispersion management in ultra-high Q microresonators. (2025) doi:10.1063/5.0028839.

36. Cardenas, J. *et al.* High Q SiC microresonators. **21**, 16882–16887 (2013).

37. Dong, X., Spiess, C., Bucklew, V. G. & Renninger, W. H. Chirped-pulsed Kerr solitons in the Lugiato-Lefever equation with spectral filtering. *Phys. Rev. Res.* **3**, (2021).

38. Wise, F. W., Chong, A. & Renninger, W. H. High-energy femtosecond fiber lasers based on pulse propagation at normal dispersion. **73**, 58–73 (2008).

39. Erdogan, T. Fiber grating spectra. *J. Light. Technol.* **15**, 1277–1294 (1997).

40. Xiao, Y., Deng, P., Zhao, Y., Yang, S. & Li, B. Three-photon excited fluorescence imaging in neuroscience : From principles to applications. 1–21 (2023) doi:10.3389/fnins.2023.1085682.

41. Xu, C., Zipfel, W., Shear, J. B., Williams, R. M. & Webb, W. W. Multiphoton fluorescence excitation : New spectral windows for biological nonlinear microscopy. **93**, 10763–10768 (1996).

42. Riemensberger, J. *et al.* Massively parallel coherent laser ranging using a soliton microcomb. *Nature* **581**, (2020).

43. He, J., Huang, D., Dai, D. & Shi, Y. Single soliton microcomb combined with optical phased array for parallel FMCW LiDAR. *Nat. Commun.* (2025) doi:10.1038/s41467-025-56483-9.

44. Geng, Y. *et al.* Wavelength-division multiplexing communications using integrated soliton microcomb laser source. **47**, 6129–6132 (2022).

45. Liu, Y. *et al.* Parallel wavelength-division-multiplexed signal transmission and dispersion compensation enabled by soliton microcombs and microrings. 1–12 (2024) doi:10.1038/s41467-024-47904-2.

46. Tienne, P. E. *et al.* Ultralow-loss tightly confining Si 3 N 4 waveguides and high- Q microresonators. **27**, 30726–30740 (2019).

47. Ippenberg, T. O. J. K. Ultra-smooth silicon nitride waveguides based on the Damascene reflow process : fabrication and loss origins. **5**, (2018).

48. Shao, Z. *et al.* Ultra-low temperature silicon nitride photonic integration platform. **24**, 1865–1872 (2016).

49. Ingchen, X. J. I. *et al.* Ultra-low-loss on-chip resonators with sub-milliwatt parametric oscillation threshold. **4**, (2017).

50. Archetti, R. I. M., Arroll, L. E. E. C., Radkowski, K. A. G. & Inzioni, P. A. M. Coupling strategies for silicon photonics integrated chips [ Invited ]. **7**, (2019).

51. Auriyal, J. U. N. Fiber-to-chip fusion splicing for low-loss photonic packaging. **6**, 3–6 (2019).

52. Kumar, S., Nauriyal, J. & Cardenas, J. Low-Loss , High Extinction Ratio Fiber to Chip Connection via Laser Fusion for Polarization Maintaining Fibers. *2023 Opt. Fiber Commun. Conf. Exhib.* 1–3 (2023) doi:10.23919/OFC49934.2023.10117121.